\documentclass[a4paper,preprint,aps]{revtex4}

\usepackage{graphicx}
\usepackage{dcolumn}
\usepackage{bm}

\newcommand{\eqa}{\begin{equation}}
\newcommand{\eqz}{\end{equation}}
\newcommand{\eqma}{\begin{eqnarray}}
\newcommand{\eqmz}{\end{eqnarray}}

\begin{document}
\newcommand{\e}{{\em e}~}
\title{The lowest singlet-triplet excitation energy of BN: a converged coupled cluster perspective}
\author{Amir Karton and Jan M. L. Martin}
\affiliation{Department of Organic Chemistry, 
Weizmann Institute of Science, 
IL-76100 Re\d{h}ovot, Israel}
\email{comartin@wicc.weizmann.ac.il}
\date{Draft version \today}
\date{{\em Journal of Chemical Physics} MS\# {\bf N6.07.009}; Received July 3, 2006; Revised August 14, 2006}

\begin{abstract}
The notoriously small $X~^3\Pi-a~^1\Sigma^+$ excitation energy of the BN diatomic has been calculated using high-order coupled cluster methods. Convergence has been established in both the 1-particle basis set and the coupled cluster expansion. Explicit inclusion of connected quadruple excitations $\hat{T}_4$ is required for even semiquantitative agreement with the limit value, while connected quintuple excitations $\hat{T}_5$ still have an effect of about 60 cm$^{-1}$. Still higher excitations only account for about 10 cm$^{-1}$. Inclusion of inner-shell correlation further reduces $T_e$ by about 60 cm$^{-1}$ at the CCSDT, and 85 cm$^{-1}$ at the CCSDTQ level.
Our best estimate, $T_e$=183$\pm$40 cm$^{-1}$, is in excellent agreement with earlier calculations and experiment, albeit with
a smaller (and conservative) uncertainty.
The dissociation energy of BN($X~^3\Pi$) is $D_e$=105.74$\pm$0.16 kcal/mol and $D_0$=103.57$\pm$0.16 kcal/mol.
\end{abstract}
\maketitle
\section{Introduction}

The lowest electronic excitation energy of the boron nitride diatomic is among the most vexing problems in small-molecule computational chemistry. Not only are the $X~^3\Pi$ and $a~^1\Sigma^+$ states nearly degenerate, but the combination of moderate multireference character in the $X~^3\Pi$ state and pathological multireference character in the $a~^1\Sigma^+$ state makes the transition energy $T_e$ excessively sensitive to the electron correlation treatment.

Martin et al.\cite{Mar92bn}, using multireference average coupled pair functional (ACPF) techniques\cite{Gda88}, found the $^3\Pi$ state to be the ground state and predicted $T_e$=381$\pm$100 cm$^{-1}$. These authors also found that (nowadays) commonly used coupled cluster methods such as CCSD(T)\cite{Rag89} yield qualitatively incorrect answers. In an MRD-CI study,  predicted . Elaborate multireference calculations by 
Mawhinney, Bruna, and Grein (MBG)\cite{Maw93}, by Peterson\cite{Pet95bn}, and by Bauschlicher and Partridge (BP)\cite{Bau96bn} obtained considerably lower $T_e$ values of 241$\pm$160 cm$^{-1}$, 190$\pm$100 and 180$\pm$110 cm$^{-1}$, respectively.
Watts\cite{Wat02bn}, at the CCSDT (coupled cluster with all single, double, and triple excitations\cite{ccsdt}) level with a cc-pVQZ (correlation consistent polarized quadruple zeta\cite{Dun89}) basis set, found $T_e$=844 cm$^{-1}$, and conjectured that this serious overestimate was due to neglect of connected quadruple ($\hat{T}_4$) and higher excitations. Both Boese et al.\cite{w3} and Tajti et al.\cite{HEAT}, in the context of high-accuracy computational chemistry protocols developed in their papers, found that, in strongly multireference systems, $\hat{T}_4$ can easily make energetic contributions on the order of the difference between the CCSDT and multireference values. (Denis\cite{Den04} crudely estimated the effect of $\hat{T}_4$ by assuming error cancellation with higher-order $\hat{T}_3$ in the singlet but not the triplet state, and predicted $T_e$=175 cm$^{-1}$.) Finally, a very recent quantum Monte Carlo (QMC) study by Lu\cite{Lu2005} in the present journal found 178(83) cm$^{-1}$, where the uncertainty band represents one standard deviation in the QMC approach.

The two most reliable experimental estimates are the noble gas matrix IR measurements of Bondybey and coworkers\cite{Bon96}, 15--182 cm$^{-1}$, and the negative ion time-of-flight photoelectron spectroscopy value of Neumark and coworkers\cite{Neu98}, 158$\pm$36 cm$^{-1}$. 

The purpose of the present work is to establish whether a converged result can be obtained at all from single-reference coupled cluster methods, whether this estimate is in agreement with the other theoretical approaches and experiment, and finally what is the breakdown of various contributions in the cluster expansion.

\section{Computational Details}

All calculations were carried out using the general coupled cluster code MRCC of K\'allay and coworkers\cite{mrcc}. The Austin-Mainz version of ACES II\cite{aces2de} was used to generate the required integrals and molecular orbitals. Unless otherwise noted, the CCSDT/cc-pVQZ reference geometries of Watts\cite{Wat02bn} were used, $r_e$($X~^3\Pi$)=1.3302 and $r_e$($a~^1\Sigma^+$)=1.2769 \AA. 

Correlation consistent\cite{Dun89} (cc-pV$n$Z), augmented correlation consistent\cite{Ken92} (aug-cc-pV$n$Z), and 
core-valence correlation consistent\cite{Woo95} (cc-pCV$n$Z) basis sets were used throughout. The largest such basis sets used, cc-pV5Z, is of $[6s5p4d3f2g1h]$ quality. Where appropriate, contributions were extrapolated to the 1-particle basis set limit using the $A+B/L^3$ formula of Halkier et al.\cite{Hal98}.

\section{Results and discussion}

All computed values are given in Table I, compared with available experimental data.

As expected, the CCSD results are grossly biased towards the triplet state ($T_e$=4432 cm$^{-1}$ at the basis set limit).
Inclusion of $\hat{T}_3$ (connected triple excitations) is required for an even qualitatively correct result, although even the CCSDT basis set limit $T_e$=827 cm$^{-1}$ is 3--4 times too large. Quasiperturbative $\hat{T}_3$ corrections such as CCSD(T) overcorrect, and wrongly predict a singlet ground state\cite{Mar92bn}. We conclude that CCSDT is the lowest acceptable level of theory for the reference geometry.
Comparison of the CCSDT/cc-pVQZ and CCSDT/cc-pV5Z values suggests that the latter is converged to within 2--3 cm$^{-1}$ with respect to the basis set.

Inclusion of $\hat{T}_4$ (connected quadruple excitation) proved essential for anything approaching quantitative accuracy. The CCSDTQ/cc-pVQZ basis set 
calculations reported here involve 419 and 391 million amplitudes, respectively, for the singlet and triplet states. They ran for two weeks each on single AMD Opteron 846 processors. At the basis set limit, $\hat{T}_4$ reduces the transition energy by 514 cm$^{-1}$. We do note --- as we have previously noted\cite{w3} for other strongly multireference systems like C$_2$($X~^1\Sigma^+_g$) --- that basis set convergence for the $\hat{T}_4$ contribution is fairly slow (unlike for systems dominated by a single reference determinant\cite{w3}). This can be rationalized in terms of very prominent double excitations in the singlet wavefunction: dynamical correlation relative to them will be dominated by double excitations, which represent quadruple excitations relative to the reference determinant.
At the CCSDTQ basis set limit, we obtain $T_e=313$ cm$^{-1}$, in agreement with Ref.\cite{Mar92bn} but still considerably higher than the other results.

Connected quintuple excitations ($\hat{T}_5$) still reduce the excitation energy by about 70 cm$^{-1}$. Comparison of the cc-pVDZ and cc-pVTZ results for this contribution suggests that it converges quite rapidly with the basis set.

Connected sextuple excitations only affect $T_e$ by --8 cm$^{-1}$, while the contribution of still higher excitations was found to be negligible. Our best estimate for the valence-only FCI basis set limit is therefore $T_e$=243$\pm$28 cm$^{-1}$, where our error bar is the sum of all the amounts covered by extrapolations.

Somewhat surprisingly, the effect of core-valence correlation is found to account for the discrepancy with earlier theoretical studies and experiment. At the CCSDT level, it lowers $T_e$ by 59$\pm$7 cm$^{-1}$, while the differential $\hat{T}_4$ core-valence contribution reduces $T_e$ by an additional 26$\pm$4 cm$^{-1}$ at the CCSDTQ level. (The all-electron CCSDTQ/cc-pCVTZ calculations, at 1 billion amplitudes each, took about one day per iteration running OpenMP-parallel on four AMD Opteron 846 CPUs. Sub-microhartree convergence requires about twenty iterations. Our attempts to carry out CCSDT/cc-pCVQZ calculations met with failure for the triplet state. Because of the clearly erratic basis set convergence behavior of the CCSD(T) energy in this case, we have chosen not to use the larger basis set data at this level of theory.)

Our final best estimate neglecting spin-orbit splitting thus becomes $T_e$=158$\pm$40 cm$^{-1}$, in excellent agreement with the earlier calculations (which likewise neglect spin-orbit splitting, it being almost an order of magnitude smaller than their stated uncertainties). Our error bar is probably somewhat conservative, as it assumes that no cancellation at all would occur between extrapolation errors in individual contributions.

The spin-orbit coupling constant of the $X~^3\Pi$ state is calculated as -24.27 cm$^{-1}$ at the CISD/cc-pVQZ (uncontracted, no $g$ functions) level using MOLPRO\cite{molpro}, in excellent agreement with the experimental value\cite{Bre85} of -25.14 cm$^{-1}$.
Its inclusion pushes up both the present calculated value and all the earlier theoretical values by these amounts: our final best estimate thus becomes $T_e$=183$\pm$40 cm$^{-1}$. This agrees with the experimental value of Neumark and coworkers\cite{Neu98} to within the respective uncertainties, and finds itself near the upper edge of the interval given by Bondybey and coworkers\cite{Bon96}. 

Finally, as a byproduct of this study, we obtain the dissociation energy of BN($X~^3\Pi$) using W4 theory\cite{w4} as $D_e$=105.74$\pm$0.16 kcal/mol and $D_0$=103.57$\pm$0.16 kcal/mol (the uncertainty being a 95\% confidence interval). This is somewhat higher than previous calculated $D_e$ values of 105.2 kcal/mol\cite{Mar92bn} and 104.2 kcal/mol\cite{Pet95bn}. The zero-point vibrational energy (ZPVE) of 2.17 kcal/mol was obtained by combining the accurate $\omega_e$ and $\omega_ex_e$ for the singlet state from Ref.\cite{Ram96} with the state difference in ZPVE from Ref.\cite{Bon96}. In Ref.\cite{w4}, \%TAE[(T)], the
percentage of the total atomization energy resulting from (T), was proposed 
as an indicator for the importance of nondynamical correlation effects.
We note that  \%TAE[(T)]=6.03\% for the $X~^3\Pi$ state (on the low end of moderate nondynamical correlation), compared to
no less than 18.63\% for the $a~^1\Sigma^+$ state (among the most severe cases surveyed in Ref.\cite{w4}).

\section{Conclusions}

Summing up, 
the notoriously small $X~^3\Pi-a~^1\Sigma^+$ excitation energy of the BN diatomic has been calculated using high-order coupled cluster methods. Convergence has been established in both the 1-particle basis set and the coupled cluster expansion. Explicit inclusion of connected quadruple excitations $\hat{T}_4$ is required for even semiquantitative agreement with the limit value, while connected quintuple excitations $\hat{T}_5$ still have an effect of about 60 cm$^{-1}$. Still higher excitations only account for about 10 cm$^{-1}$. Inclusion of inner-shell correlation further reduces $T_e$ by about 60 cm$^{-1}$ at the CCSDT, and 85 cm$^{-1}$ at the CCSDTQ level.
Our best estimate, $T_e$=183$\pm$40 cm$^{-1}$, is in excellent agreement with earlier calculations and experiment, albeit with
a smaller (and conservative) uncertainty.
The dissociation energy of BN($X~^3\Pi$) is $D_e$=105.74$\pm$0.16 kcal/mol and $D_0$=103.57$\pm$0.16 kcal/mol.

\acknowledgments

Research was supported by the Israel Science Foundation (grant 709/05), the Minerva Foundation (Munich, Germany), and the Helen and Martin Kimmel Center for Molecular Design. JMLM is the incumbent of the Baroness Thatcher Professorial Chair of Chemistry and a member {\em ad personam} of the Lise Meitner-Minerva Center for Computational Quantum Chemistry. The authors thank Dr. Mih\'aly K\'allay for kind assistance with MRCC and access to a prerelease version of the code, and Prof. John D. Watts (Jackson State University) for helpful correspondence.

\clearpage
\squeezetable
\begin{table}
\caption{$X~~^3\Pi-a^1\Sigma^+$ transition energy (cm$^{-1}$)}
\begin{tabular}{lrrrrrrrr}
\hline\hline
valence correlation \\
\hline
                  & cc-pVDZ & aug-cc-pVDZ & cc-pVTZ & aug-cc-pVTZ$^a$ & cc-pVQZ & cc-pV5Z & Best estimate & Running total\\
\hline
CCSD              & 4250.7  &  4619.8     & 4375.6  &    4469.7   & 4420.7  & 4427.4  & 4432.2        & --- \\
CCSD(T)           & -141.2  &             & -180.7  &             & -175.4  & -181.1  & -199.9$^b$    & ---\\
CCSDT             &  814.6  &  1203.0     &  826.3  &     931.7   &  831.6  &  829.6  &  827.4        & 827.4 \\
CCSDTQ$-$CCSDT      & -323.7  &  -375.3     & -466.6  &    -477.9   & -494.2  &  ---    & -514.4        & 313.0 \\
CCSDTQ5$-$CCSDTQ    &  -50.6  &   -53.9     &  -58.3    &     ---     &  ---    &  ---    &  -61.2        & 251.8 \\
CCSDTQ56$-$CCSDTQ5  &   -7.6  &   ---       &  ---    &     ---     &  ---    &  ---    &   -7.6        & 244.2 \\
FCI$-$CCSDTQ56      &   -0.9  &   ---       &  ---    &     ---     &  ---    &  ---    &   -0.9        & 243.3 \\
\hline
inner shell corr.\\
\hline
                  &cc-pCVDZ & cc-pCVTZ    & cc-pCVQZ & cc-pCV5Z &         &         & Best estimate & Running total\\
CCSD(T)           &  -15.2  &  -15.7      &  -6.2    &  +4.8       &         &    &  +16.2          &  ---\\
CCSDT             &  -36.0  &  -52.4      &          &             &         &         &  -59.3          & 184.0\\
CCSDTQ-CCSDT      &  -12.1  &  -21.6          &          &             &         &         &  -25.6          & 158.4\\
\hline          
Best estimate, this work&&&&&&&&158$\pm$40$^d$\\
Incl. spin-orbit$^c$&&&&&&&&183$\pm$40\\
MRACPF, Martin et al.\cite{Mar92bn}     &&&&&&&&381$\pm$100$^d$\\
MRDCI, Mawhinney et al.\cite{Maw93}     &&&&&&&&241$\pm$160$^d$\\
ICMRCI, Peterson\cite{Pet95bn}     &&&&&&&&190$\pm$100$^d$\\
ICMRCI, BP\cite{Bau96bn}     &&&&&&&&180$\pm$110$^d$\\
QMC, Lu\cite{Lu2005} &&&&&&&&178$\pm$83$^d$\\
Expt.(matrix)\cite{Bon96}     &&&&&&&&15--182\\
Expt.(gas phase)\cite{Neu98}  &&&&&&&&158$\pm$36$^e$\\
\hline\hline
\end{tabular}
\begin{flushleft}

(a) cc-pVTZ basis set used on boron.\\

(b) extrapolated from CCSD(T)/cc-pV5Z value and -189.1 cm$^{-1}$ at the CCSD(T)/cc-pV6Z level.\\

(c) Expt. $A_0$=-25.14 cm$^{-1}$\cite{Bre85}, calc. $A_e$=-24.3 cm$^{-1}$ (this work).\\

(d) Value does not include spin-orbit splitting in triplet state.\\

(e) From observed $T_0$=0.031$\pm$0.004 eV\cite{Neu98} and ZPVE difference from Ref.\cite{Bon96}, assuming 4~cm$^{-1}$ uncertainty on ZPVE difference.
\end{flushleft}

\end{table}

\end{document}